\documentclass[
]{ceurart}

\usepackage{listings}
\usepackage{url}
\usepackage{booktabs}
\usepackage[linesnumbered,ruled,vlined]{algorithm2e}
\usepackage{wrapfig}

\begin{document}

\newcommand{\vct}[1]{\boldsymbol{#1}}
\newcommand{\mat}[1]{\boldsymbol{#1}}
\newcommand{\mypar}[1]{\noindent\textbf{#1}}
\newcommand{\mysubpar}[1]{\noindent\underline{\textit{#1}}}
\renewcommand{\sectionautorefname}{Sect.}
\renewcommand{\subsectionautorefname}{Sect.}
\renewcommand{\subsectionautorefname}{Sect.}
\renewcommand{\tableautorefname}{Table}
\renewcommand{\algorithmautorefname}{Algorithm}

\copyrightyear{2025}
\copyrightclause{Copyright for this paper by its authors.
  Use permitted under Creative Commons License Attribution 4.0
  International (CC BY 4.0).}

\conference{Ital-IA 2025: 5th National Conference on Artificial Intelligence, organized by CINI, June 23-24, 2025, Trieste, Italy}

\title{Empirical Quantification of Spurious Correlations in Malware Detection}

%\tnotetext[1]{You can use this document as the template for preparing your publication. We recommend using the latest version of the ceurart style.}

\author[1]{Bianca Perasso}[%
email=bianca.perasso@gmail.com,
]

\author[1]{Ludovico Lozza}[%
email=ludovico.lozza@hotmail.com,
]

\author[1]{Andrea Ponte}[%
orcid=0009-0001-5647-1077,
email=andrea.ponte@edu.unige.it,
]
\cormark[1]

\author[1]{Luca Demetrio}[%
orcid=0000-0001-5104-1476,
email=luca.demetrio@unige.it,
]
\cormark[1]

\author[1]{Luca Oneto}[%
orcid=0000-0002-8445-395X,
email=luca.oneto@unige.it,
]

\author[1,2]{Fabio Roli}[%
orcid= 0000-0003-4103-9190,
email=fabio.roli@unige.it,
]

\address[1]{Università degli Studi di Genova, Italy}
\address[2]{Università degli Studi di Cagliari, Italy}
\cortext[1]{Corresponding author.}
%\fntext[1]{These authors contributed equally.}

\begin{abstract}
  End-to-end deep learning exhibits unmatched performance for detecting malware, but such an achievement is reached by exploiting spurious correlations -- features with high relevance at inference time, but known to be useless through domain knowledge.
  While previous work highlighted that deep networks mainly focus on metadata, none investigated the phenomenon further, without quantifying their impact on the decision. 
  In this work, we deepen our understanding of how spurious correlation affects deep learning for malware detection by highlighting how much models rely on empty spaces left by the compiler, which diminishes the relevance of the compiled code.
  Through our seminal analysis on a small-scale balanced dataset, we introduce a ranking of two end-to-end models to better understand which is more suitable to be put in production.
\end{abstract}


\maketitle
\section{Introduction}
Antivirus programs are now deeply integrated with machine learning components, extending the defensive capabilities of endpoints.
In particular, one prominent approach is posed by end-to-end techniques, which directly digest the raw bytes of programs, thus learning an intermediate representation directly from data.
This is opposed to regular feature extraction processes, that pre-process each sample to extract aggregated and compressed information, later used at training time.
While being time-consuming~\cite{anderson2018ember}, feature extraction in cybersecurity contexts are easily prone to pre-processing errors~\cite{ponte2025slifer}, since malicious actors always try to complicate their programs to impede the analysis.
Hence, even if the requirement of available samples is higher, end-to-end models ditch this problem entirely, still exhibiting excellent performance in production~\cite {raff2018malware, krvcal2018deep, coull2019activation,gibert2020rise}.
However, as a downside, these end-to-end models are completely opaque in terms of which are the relevant features used to compute predictions. 
In particular, previous work~\cite{demetrio2019explaining, coull2019activation, bose2020explaining} debated whether Windows malware detectors implemented with deep neural networks are affected by \emph{spurious correlations} -- correlations between the predicted class and features that are known to be not relevant in terms of domain knowledge.
While they all acknowledge an excessive focus on the metadata of input samples, none of the previous work either (i) tried to quantify how much these models rely on spurious correlations, and (ii) ignore other spurious correlations that might be learned by those models.
Hence, in this preliminary analysis, we investigate and propose an empirical quantification of three spurious correlations that can be learned by Windows malware detectors, while also highlighting the relevance of features known to be important.
Such is achieved by leveraging \emph{integrated gradients}~\cite{sundararajan2017axiomatic} (IG), a gradient-based method for computing the relevance of each input feature of deep learning models.
Relying on domain knowledge, we select the locations inside samples where there might be spurious correlations, and we estimate their relevance through the $\ell_2$ norm.
Such aggregated relevance is then normalized to gain an understanding of their impact in the decision.
Through an experimental analysis on a balanced small-scale dataset, we show how two state-of-the-art end-to-end deep networks are affected by spurious correlations, impacting the relevance attributed to the compiled code which should be, in theory, the most relevant portion of any executable.

\begin{figure}[t]
    \centering
    \includegraphics[width=\linewidth]{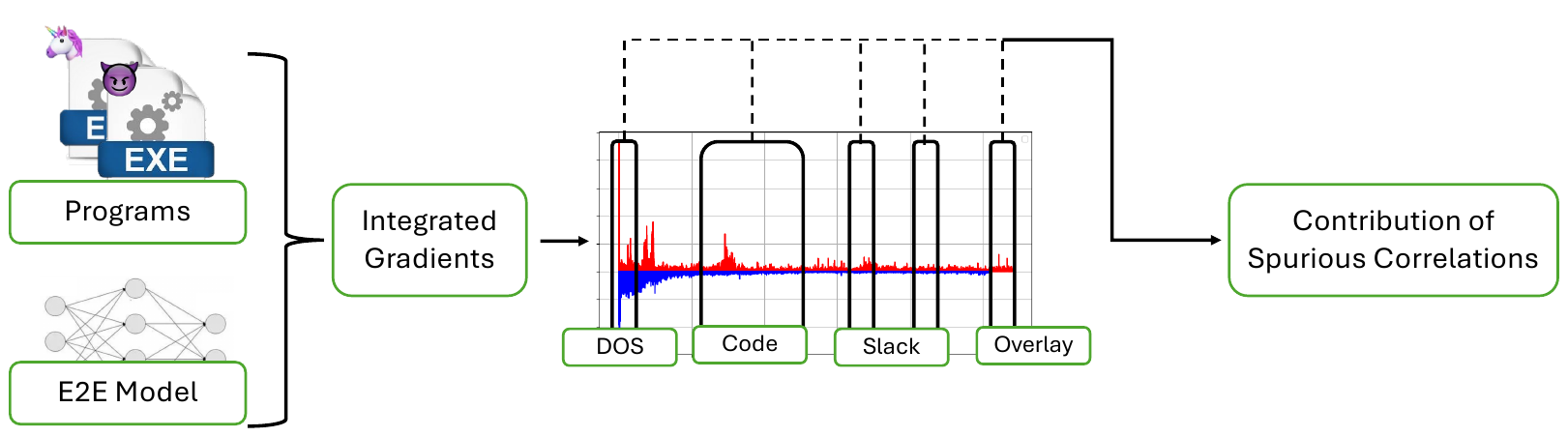}
    \caption{Recap of our analysis: given an end-to-end model and Windows programs, we compute Integrated Gradients, and we extract the attributed relevance of (i) the DOS header, (ii) the slack bytes, (iii) the overlay, and (iv) the ``.text'' section. These are aggregated to obtain a score, which hint how much the input model relies on spurious correlations when computing predictions.}
    \label{fig:abs}  
\end{figure}

\section{Background and Related Work}\label{sec:background}
Before describing our research work, we briefly present the key concepts to fully understand our manuscript. In the following, we explain the type of data we deal with, how a malicious sample can be detected, and how this decision can be explained. 

%\subsection{Background}
\begin{wrapfigure}{r}{0.6\textwidth}
    \centering 
    \includegraphics[width=\linewidth]{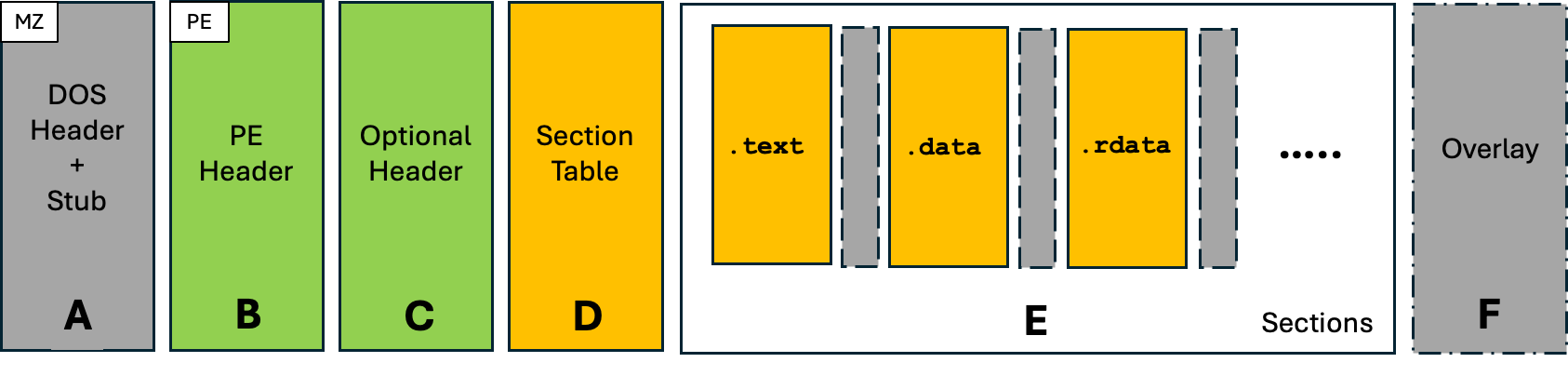}
    \caption{Representation of the Windows PE File Format.}
    \label{fig:pe}
\end{wrapfigure}
\mypar{Windows PE File Format.} In this work, we focus on malware detection for Windows operating systems. Each Windows program is stored in Portable Executable (PE) format\footnote{\label{format}\url{https://learn.microsoft.com/en-us/windows/win32/debug/pe-format}}. 
The PE format is composed of several parts (\autoref{fig:pe}), containing everything the OS needs to store, load, and execute the program: (\textbf{A}) the DOS header and stub, which compose a valid DOS program, are components kept only for backward compatibility purposes, and they are not informative about the program structure;  (\textbf{B}) the PE header is the real header of the PE format, and, together with the (\textbf{C}) Optional Header, contains essential information for storing and loading the program; the Section Table (\textbf{D}) and (\textbf{E}) the Sections, that instructs the OS where to find the real content of the program, such as \textit{.text}, being the compiled code of the program (for all the others, we refer to the documentation of the format).
While being necessary for each program to comply with this structure, the format itself permits modifications. These are often named code caves, and together with backward compatibility parts (\textbf{A}), leave an open field to malware developers~\cite{anderson2017evading, demetrio2021functionality}. For instance, due to alignment requirements, the compiler allocates more space than needed to store sections: these uninitialized bytes between sections are called \textit{slack bytes}, represented in gray in \autoref{fig:pe} (\textbf{E}).
Moreover, the PE format does not forbid the appending of bytes at the end of the last section (\textit{overlay space}, depicted in \textbf{F}).

\mypar{End-to-end Malware Detectors.} In recent years, malware detection methods have fully embraced the AI paradigm, training Machine Learning (ML) and Deep Learning (DL) models on past data~\cite{gibert2020rise}. Models can learn from different kinds of malware analysis, which can extrapolate knowledge from the sole program structure (static analysis) or the behavior of malicious samples, previously executed in isolated environments (dynamic analysis). 
In this work, we leverage end-to-end static detectors, which do not rely on feature extraction, which can be faulty~\cite{ponte2025slifer}, but can learn directly from raw bytes, as proposed by Raff et al.~\cite{raff2018malware}, Coull et al.~\cite{coull2019activation} and Kr{\v{c}}{\'a}l et al~\cite{krvcal2018deep}. Generally, these DL architectures treat each PE byte as an input for the network, which is represented as a vector using an embedding layer. This approach produces matrices in which each row corresponds to a byte, represented as a vector generated by the embedding layer. After embedding, the input matrices pass through a series of convolutional and pooling layers, followed by fully connected layers that generate the final output probability.

\mypar{Integrated Gradie nts.} Explaining predictions of deep neural networks is a thorny challenge, due to the black-box nature of these complex models. 
In their work, Sundararajan et al.~\cite{sundararajan2017axiomatic} propose \textit{Integrated Gradients}, an attribution method that compute the relevance of each input feature in computing the prediction of input samples.
These contributions are computed with respect to a \textit{baseline}, representing a null signal from which the relevant features ``emerge''. 
To calculate them, the method accumulates the gradients of the model to interpret with respect to the input along a straight-line path from the baseline to the actual input, and then it averages all them to find the relevance of each feature.

\mypar{Related Work on Spurious Correlations in Malware Detection.}  Recent work rise the debate the presence of \textit{spurious correlation} in malware detection~\cite{demetrio2019explaining, coull2019activation, bose2020explaining, arp2022and}. 
While these work analyses peculiar behaviors of malware detectors, supporting both the presence of useless features that might cause drop in robustness~\cite{demetrio2019explaining, bose2020explaining}, or deep networks being able to learn relevant locations inside headers~\cite{coull2019activation}, none of them proposes an empirical way to quantify these awkward phenomena.

\section{Measuring Spurious Correlations for Malware Detectors}
\begin{wrapfigure}{r}{0.5\textwidth}
    \begin{minipage}{\linewidth}
        \begin{algorithm}[H]
        \SetAlgoLined
        \KwData{$f$, a malware detector; $\mat X$, a set of Windows program}
        \KwResult{$\vct r_f$, reliance on spurious correlation}
        \DontPrintSemicolon
        $\vct{r_{dos}},\vct{r_{slack}}, \vct{r_{ovelay}}, \vct{r_{.text}} \leftarrow 0$\;\label{line:init}
        \For{$\vct x \in \mat X$}{\label{line:forsample}
            $\vct{ig} \leftarrow IG(f, \vct x)$\;\label{line:ig}
            $\eta \leftarrow 1 / (|\mat X| ||\vct{ig}||_2^2)$\;
            $\vct{r_{dos}}  \leftarrow \vct{r_{dos}} + \eta||\text{select\_dos}(\vct{ig})||_2^2$\;\label{line:dos}
            $\vct{r_{slack}} \leftarrow  \vct{r_{slack}} + \eta||\text{select\_slack}(\vct{ig})||_2^2$\;\label{line:slack}
            $\vct{r_{ov}} \leftarrow \vct{r_{ov}} + \eta||\text{select\_overlay}(\vct{ig})||_2^2$\;\label{line:overlay}
            $\vct{r_{.text}} \leftarrow \vct{r_{.text}} + \eta||\text{select\_text}(\vct{ig})||_2^2$\;\label{line:text}
            
        }
        \textbf{return} $\vct{r_{.text}} - (\vct{r_{dos}} + \vct{r_{slack}} + \vct{r_{ovelay}})$\;\label{line:return}
        \caption{Impact of Spurious Correlation in Windows malware detectors}
        \label{algo:rank}
    \end{algorithm}
\end{minipage}
\end{wrapfigure}
We now illustrate how we quantify the relevance attributed by end-to-end Windows malware detectors to known spurious correlations, and we report our methodology in \autoref{algo:rank}.
First, we leverage domain knowledge on Windows PE file format to isolate parts of each executable file that are known to be semantically irrelevant for malware detection.
While there might be plenty, in this paper we focus on three specific parts that should not be relevant at inference time: the \textit{DOS header}, the \textit{slack bytes}, and the \textit{overlay space}.
Then, given an end-to-end Windows malware detector, for each sample (\autoref{line:forsample}) we compute feature attributions using Integrated Gradients (IG) ~\cite{sundararajan2017axiomatic} (\autoref{line:ig}).
If the attributions computed by IG on the selected regions are different from zero, it means that the model is considering those to compute predictions with other features as well, thus confirming the presence of spurious correlations. 
These attribution, divided in the selected regions through domain knowledge, are characterized by their $\ell_2$ norm (\autoref{line:dos} -- \autoref{line:text}), and normalized accordingly with the total $\ell_2$ norm of the attributions.
In this way, we obtain a score for each spurious correlation that has an higher bound on the total norm of the attribution: if this ratio leans towards 1, it mean that all the prediction relies on such portion of the executable.
We also select the \textit{.text section} to show how much relevance is given to the most important part of an executable, expecting it to be higher than the other numbers.
Finally, we compute a score based on the previous steps, by subtracting from the average relevance of the compiled code, the average relevance of the spurious correlations (\autoref{line:return}).
If this metric is positive, the analyzed model is attributing most of relevance to the really important features. If it is zero or negative, it means that the analyzed model is giving more relevance to spurious correlations, harming its reliability.

\section{Experimental Analysis}
%\mypar{Experimental Setup.}

\mypar{Experimental Setup.} We detail here how we have setup our analysis, describing which data and deep neural networks we have used.

\begin{table}[]
\begin{tabular}{@{}ccccccc@{}}
\toprule
\textbf{Model} & \textbf{Data} & \textbf{DOS} & \textbf{Slack} & \textbf{.text} & \textbf{Overlay} & \textbf{Aggregate Score} \\ \midrule
\multirow{2}{*}{MalConv} & goodware & 0.0493 & 0.2567 & 0.5388 & 0 & \multirow{2}{*}{0.2939} \\
                         & malware  & 0.0705    & 0.0911    & 0.5111    & 0  \\ \midrule
\multirow{2}{*}{BBDNN}   & goodware & 0.0228 & 0.4082 & 0.4802 & 0 & \multirow{2}{*}{0.2502} \\
                         & malware  & 0.0273    & 0.1117    & 0.5529    & 0  \\ \bottomrule
\end{tabular}
\caption{Magnitude of integrated gradients applied to both malware and goodware samples, characterized by the output of \autoref{algo:rank}. We also report the relevance of the four analyzed PE regions (DOS, Slack, .text, and Overlay) to clarify the contribution of each of them.}
\label{tab:aggregated_ig_results}
\end{table}

\mysubpar{Dataset.} We use a dataset composed of 210 malware samples and 210 goodware samples. We take malware samples from the Speakeasy testset~\cite{trizna2022quo}, sampling 30 PEs for each family (\textit{Backdoor}, \textit{Coinminer}, \textit{Dropper}, \textit{Keylogger}, \textit{Ransomware}, \textit{RAT} and \textit{Trojan}).
We take benign samples from a fresh installation of Windows 11, inside \textit{sys32} folder. 

\mysubpar{Models.} In our experimental work, we leverage two CNN architectures, mentioned in \autoref{sec:background}: the first is MalConv, proposed by Raff et al.~\cite{raff2018malware}, the second is BBDNN, proposed by Coull et al.~\cite{coull2019activation}. MalConv embeds input sequences (truncated to 1 MB) with an 8-dimensional embedding space, then it implements a single gated convolutional layer~\cite{dauphin2017language} with global max-pooling, followed by a single fully-connected layer and softmax. BBDNN is quite similar to MalConv, but it has a larger embedding space (10 dimensions), and forwards input sequences to five alternating convolutional and pooling layers and finally to a fully connected layer with a sigmoid function. BBDNN results in a deeper architecture, and it pays this price by truncating input byte sequences to 100 KB to have reasonable training times. 
Both models are trained on the EMBER dataset~\cite{anderson2018ember}, and we use the pretrained version included into \textit{maltorch}\footnote{\url{https://github.com/zangobot/maltorch}} library.

\mysubpar{Setup of Integrated Gradients.} The method requires two hyper-parameters: the baseline from which compute the attributions, and the size of the interpolation.
Regarding the baseline, we adopt the same setting provided by previous work~\cite{demetrio2019explaining}, and we set the baseline as the empty file, thus filled with the special character 256.
Thereafter, we set the interpolation size to evaluate 50 gradients.

\mysubpar{Setup of \autoref{algo:rank}.} Since many malware samples are obfuscated, thus they have renamed or removed the \texttt{.text} section, we use as code the first section that is marked as executable to conclude our analysis. In this way, we ensure that each sample is correctly treated as specified in \autoref{algo:rank}.

\mypar{Experimental Results.} We now present the results of our analysis, by computing the relevance of the four analyzed PE regions using Integrated Gradients (IG) method, applied to both MalConv and BBDNN models. 
For each model, we report the average values computed for the four analyzed regions of PE files, both for goodware and for malware, and all results are presented in \autoref{tab:aggregated_ig_results}.
Even in this simplified setting, we can observe that: 
(i) on average, both models rely more on the bytes found in the executable section, which is evidenced by the positive aggregated scores, nevertheless these are diminished by the presence of the spurious correlations; 
(ii) there is a notable distinction between goodware and malware samples, particularly the values in the Slack regions are higher for the goodware samples with respect to the malware ones, suggesting that the models might focus too much on unused space; 
(iii) comparing the two models, their aggregated scores appear largely similar on average, however in this analysis we do not account the different models' input spaces; 
(iv) for both models and across both datasets, the relevance attributed to overlay region is always zero, which may indicate that the models are not attributing relevance to those bytes. However, it is important to notice that the majority of the samples is longer than the actual file sizes: there are 283 files larger than 100 KB (BBDNN's input space) and 100 files larger than 1 MB (MalConv's input space). 
\begin{wrapfigure}{r}{0.6\textwidth}
    \centering 
    \includegraphics[width=\linewidth]{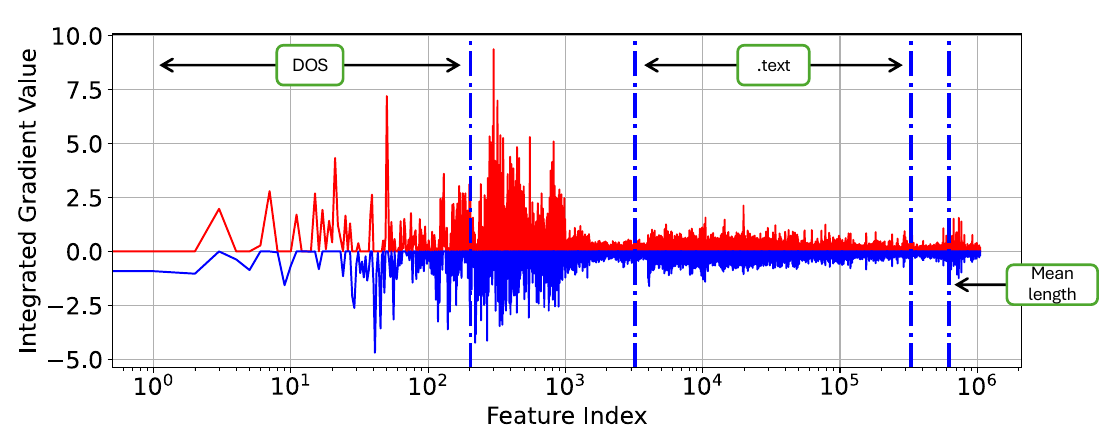}
    \caption{Average attribution of Integrated Gradients computed on MalConv. Red / Blue bars symbolize attribution towards the malicious / benign class.}
    \label{fig:all_ig}
\end{wrapfigure}
These trends are summarized by the average response of Integrated Gradients in \autoref{fig:all_ig}, where it is clear that plenty of relevance is attributed to bytes in executable sections, but there are peaks in other regions as well.
Also, this aggregated view highlight that it is very likely that many other spurious correlations have been learned by these models which are located in other places other than the ones we have analyzed.

\section{Conclusions}

\mypar{Limitations and Future Work.}
Our preliminary analysis is timely, but it has some limitations.
First, we tested our methodology with a small-scale dataset using only two state-of-the-art end-to-end neural network. 
Since this analysis depends on both data and trained models, results might change accordingly.
Nevertheless, the dataset we have used is balanced, both in terms of ratio between goodware and malware, but also in terms of the malware families we have considered.
Secondly, our results do not consider the length of the input window of each network.
Hence, larger sections, which are kept without being cut, contribute more to the total score computed for the model.
However, with our methodology we are characterizing how much the spurious correlations contribute to the norm of the calculated attribution, showing that they absorb a consistent fraction of it.
Thus, as future work, we will consider larger dataset like analyzing the whole Speakeasy dataset~\cite{trizna2022quo}, and other state-of-the-art data sources, also including other end-to-end neural networks for malware detection~\cite{krvcal2018deep}.
Lastly, while we only covered spurious correlations on static analysis, we will extend our work to also cover spurious correlation in \emph{dynamic analysis}, which track the execution of programs and train models on summary reports of their activity.

\mypar{Final Remarks.}
In this preliminary analysis, we proposed a methodology for quantifying the impact of spurious correlations in end-to-end Windows malware detectors.
While most of the prediction is focused inside the section containing code, a consistent amount of the attribution is also given to spurious correlation.
Through our analysis, we introduce a way to characterize such reliance to spurious correlations in malware detectors, paving the road towards novel techniques that can, in the near future, serve as the first benchmark of these technologies, providing a more reliable way to deploy those.

\section*{Acknowledgments}
The authors acknowledge the help of Daniel Gibert for releasing trained models in \textit{maltorch} library. 
This work was partially supported by projects SERICS (PE00000014); FAIR (PE00000013) under the NRRP MUR program funded by the EU - NGEU, and FISA-2023-00128 funded by the MUR program “Fondo italiano per le scienze applicate”.

\bibliography{sample-ceur}

\end{document}